\documentclass[preprint,showpacs,preprintnumbers,amsmath,amssymb]{revtex4}%
\usepackage{graphicx}
\usepackage{dcolumn}
\usepackage{bm}
\usepackage{amsmath}
\usepackage{epstopdf}
\usepackage{amsfonts}
\usepackage{amssymb}%
\providecommand{\U}[1]{\protect\rule{.1in}{.1in}}
\newcommand{\be}{\begin{equation}}
\newcommand{\en}{\end{equation}}
\newcommand{\bea}{\begin{eqnarray}}
\newcommand{\ena}{\end{eqnarray}}
\begin{document}
\title{Dynamics of warm power-law plateau inflation with a generalized inflaton decay rate: predicctions and constraints after Planck 2015}
\author{Abdul Jawad}
\email{abduljawad@ciitlahore.edu.pk; jawadab181@yahoo.com}
\affiliation{Department of Mathematics, COMSATS Institute of Information Technology, Lahore-54000, Pakistan.}
\author{Nelson Videla}
\email{nelson.videla@ing.uchile.cl}
\affiliation{Departamento de F\'{\i}sica, FCFM, Universidad de Chile, Blanco Encalada 2008, Santiago, Chile}
\author{Faiza Gulshan}
\email{fazi.gull@yahoo.com}
\affiliation{Department of Mathematics, Lahore Leads University,\\
Lahore-54000, Pakistan.}

\date{\today}

\begin{abstract}

In the present work we study the consequences of considering 
a new family of single-field inflation models, called power-law plateau inflation, in
the warm inflation framework. We consider 
the inflationary expansion is driven by a standard scalar field with a
decay ratio $\Gamma$ having a generic power-law dependence with the scalar field $\phi$ and the 
temperature of the thermal bath $T$ given by $\Gamma(\phi,T)=C_{\phi}\,\frac{T^a}{\phi^{a-1}}$. Assuming that our model evolves according to the strong dissipative regime, 
we study the background and perturbative dynamics, obtaining the most relevant inflationary observables
as the scalar power spectrum, the scalar spectral index and its running, and the tensor-to-scalar ratio. The free parameters characterizing our model are constrained by considering the essential condition for warm inflation,  the conditions for the model evolves according to the strong dissipative regime, and the 2015 Planck results through the $n_s-r$ plane. For completeness,
we study the predictions in the $n_s-dn_s/d\ln k$ plane. The model 
is consistent with a strong dissipative dynamics and predicts values for the tensor-to-scalar ratio and for the running of the scalar spectral index consistent with current bounds imposed by Planck, and we conclude that the model is viable.
\end{abstract}

\maketitle



\section{Introduction}

The inflationary universe has become in the most acceptable framework in describing the physics of the very early universe. Besides of solving most of the shortcomings of the hot big-bang scenario, like the horizon, the flatness, and the
monopole problems \cite{Starobinsky:1980te,R1,R106,R103,R104,R105,Linde:1983gd}, inflation also generates a mechanism to explain the large-scale structure (LSS) of the universe  \cite{Starobinsky:1979ty,R2,R202,R203,R204,R205}
and the origin of the anisotropies observed in the cosmic microwave background (CMB) radiation \cite{astro,astro2,astro202,Hinshaw:2012aka,Ade:2013zuv,Ade:2013uln,Ade:2015xua,Ade:2015lrj}, since primordial density perturbations may be sourced from quantum fluctuations of the inlaton scalar field during the inflationary expansion.
The standard cold inflation scenario is divided into two regimes: the slow-roll and reheating phases. In the slow-roll period the universe undergoes an accelerated expansion and all interactions
of the inflaton scalar field with other field degrees of freedom are typically neglected. Subsequently, a
reheating period \cite{Kofman:1994rk, Kofman:1997yn,Allahverdi:2010xz,Amin:2014eta} is invoked to end the brief acceleration. After reheating, the universe is filled
with relativistic particles and thus the universe enters in the radiation big-bang epoch. 

Upon comparison to the current cosmological and astronomical observations, specially those related with the CMB
temperature anisotropies, it is possible to constrain the inflationary models. In particular, the constraints in the $n_s-r$ plane
give us the predictions of a number of representative inflationary
potentials. Recently, the Planck
collaboration has published new data of enhanced precision of
the CMB anisotropies \cite{Ade:2015lrj} . Here, the Planck full mission
data has improved the upper bound on the tensor-to-scalar ratio
$r_{0.002} < 0.11$($95\%$ CL) which is similar to obtained from  \cite{Ade:2013uln} , in which
$r < 0.12$ ($95\%$ CL). In particular, some representative models, as chaotic inflation, which predict a large value of the tensor-to-scalar ratio $r$, are ruled out by the data. As it was reported in Ref.\cite{Martin:2013nzq}, the Planck data tends to support plateau-like inflaton scalar potentials, which are asymptotically constant. The Starobinsky $R^2$ \cite{Starobinsky:1980te} and Higgs inflation \cite{Bezrukov:2010jz} are the most representative models with this class of potentials, and more recently, the $\alpha$-attractors models \cite{Kallosh:2013yoa,Kallosh:2014rga}. For this last class of models, the approach 
to the inflationary plateau is exponential, being indistinguishable from Starobinsky and Higgs inflation models. In this direction, and
starting from global supersymmetry and considering a superpotential, it was proposed in \cite{Dimopoulos:2014boa,Dimopoulos:2015aca} a new class of models called shaft inflation. As opposed to Starobinsky and Higgs inflation, the approach to the plateau is power-law. In a in a subsequent work \cite{Dimopoulos:2016zhy},
another new family of inflationary models is studied, proposing a phenomenological potential
\begin{equation}
V(\phi)=V_0\left[\frac{\left(\frac{\phi}{M_p}\right)^n}{\left(\frac{\phi}{M_p}\right)^n+\alpha^n}\right]^q,\label{Vplateau}
\end{equation}
where $n$ and $q$ are real parameters, $V_0$ is a constant density scale and $\alpha \equiv \frac{M}{M_p}$, with $M$ being a
mass scale and $M_p=2\textup{.}43\times 10^{18}$ GeV denotes the reduced Planck mass. In addition it is required that $\phi \geq M$, or equivalently $\phi \geq \alpha M_p$ , otherwise this model
is indistinguishable from monomial inflation where $V \propto \phi^{nq}$. It was
demonstrated in the $r-n_s$ plane that the predictions of power-law plateau inflation are distinct and testable compared
to several inflation models. In particular, the case $n=2$ and $q=1$ corresponds to the best choice of model in Ref.\cite{Dimopoulos:2016zhy}.
Despite this atractiveness, in order to be a realistic model, it needs to be embedded 
in a convenient theoretical framework.

On the other hand, some classes of inflaton models excluded by current data in the standard cold inflation scenario
can be saved  in the warm inflation scenario, which is an alternative mechanism for having successful inflation. The warm inflation
scenario, as opposed to standard cold inflation, has the essential feature that a reheating phase is avoided at the end of the accelerated expansion due to the decay of the inflaton into radiation
and particles during the slow-roll phase \cite{warm1,warm2}. During warm inflation, the temperature of the universe does not drop dramatically and the universe can smoothly enter into the decelerated, radiation-dominated period, which is essential for a successful
big-bang nucleosynthesis. In the warm inflation scenario, dissipative effects are important during the accelerated expansion,
so that radiation production occurs concurrently with the accelerated expansion. For a representative list
of recent references see Refs.\cite{Bastero-Gil:2014raa,Bastero-Gil:2015nja,Panotopoulos:2015qwa,Bastero-Gil:2016qru,Visinelli:2016rhn,Gim:2016uvv,Oyvind Gron:2016zhz,Benetti:2016jhf,Peng:2016yvb,Sayar:2017pam}. The dissipative effect arises from a friction term or dissipative coefficient $\Gamma$, which describes the processes of the scalar field dissipating into a thermal bath via its interaction with other field degrees of freedom. The effectiveness of warm inflation may be parametrized by the ratio $R\equiv \Gamma/3H$. The weak dissipative regime for warm inflation is for $R\ll 1$, while for $R\gg1$, it is the strong dissipative regime for warm inflation. It is important to emphasize that the dissipative coefficient $\Gamma$ may be computed from first principles in quantum field theory
considering that $\Gamma$ encodes the microscopic physics resulting from the interactions between the inflaton and the other fields that can be present. For instance, by considering different decay mechanisms, it is possible to obtain several expressions for the dissipative coefficient $\Gamma$. In particular, in Refs.\cite{26,BasteroGil:2012cm,Bartrum:2013fia}, a supersymmetric model
containing three superfields $\Phi$, $X$, and $Y$ has been studied with a superpotential $W=f\left(\Phi\right)+\frac{g}{2}X^2+\frac{h}{2}XY^2$, where the scalar components of the superfields are $\phi=\sqrt{2}\left\langle \Phi \right\rangle$, $\chi$, and $y$, respectively. The inflaton scalar potential is given by $V(\phi)=\left|f^{\prime}(\phi)^2\right|$, which spontaneously breaks supersymmetry (SUSY). By coupling the inflaton to the bosonic and fermionic $X$ fields and their subsequent decay into $Y$ scalars and fermions, which form the thermal bath, and for the case of low-temperature regime, when the mass of the catalyst field $m_{\chi}$ is larger than the temperature $T$, the resulting dissipation coefficient can be well described by the expression $\Gamma=C_{\phi}\frac{T^3}{\phi^2}$, where $C_{\phi}$ is a dimensionless parameter related to the dissipative microscopic dynamics. For this particular case $C_{\phi}\simeq \frac{1}{4}\alpha_h N_X$, with $\alpha_h=h^2 N_Y/4\pi\lesssim 1$ and $N_{X,Y}$ denote the multiplicity of chiral superfields. In this direction, SUSY ensures that quantum and thermal corrections to the effective potential are under control \cite{Berera:2008ar}. As another example, in Ref.\cite{Berera:1998px}, it was demonstrated that a form $\Gamma=C_{\phi} \phi^2/T$, in principle with problems of large thermal corrections for the models studied in Refs.\cite{28, PRD}, may produce a consistent warm inflation scenario for a specific model. On the other hand, it was shown for the first time in Ref.\cite{Bastero-Gil:2016qru} that warm inflation can be realized by directly coupling the inflaton to a few light fields instead to consider indirect couplings to light fields through heavy mediator fields, as in Refs.\cite{26,BasteroGil:2012cm,Bartrum:2013fia}. Then, the expression obtained for $\Gamma$ turns out to be $\Gamma=C_{\phi}T$.  

Following Refs.\cite{BasteroGil:2012cm,Zhang:2009ge,BasteroGil:2010pb},
a general parametrization of the dissipative coefficient
$\Gamma(T,\phi)$ can be written as
\begin{equation}
\Gamma(T,\phi)=C_{\phi}\,\frac{T^{a}}{\phi^{a-1}}. \label{G}%
\end{equation}

This expression includes the several cases mentioned above. Specifically, for the value $a=3$, this case corresponds to a low-temperature regime, when the mass of the catalyst field $m_{\chi}$ is larger than the temperature $T$ \cite{26,BasteroGil:2012cm,Bartrum:2013fia}. On the other hand, $a=1$, i.e, $\Gamma \propto T$ corresponds to \cite{Bastero-Gil:2016qru}. For $a=0$, the dissipative coefficient represents an exponentially decaying propagator in the high-temperature regime. Finally, for $a=-1$, i.e., $\Gamma \propto \phi^2/T$ agrees with the non-SUSY case \cite{Berera:1998px,28, PRD}.

 Additionally, thermal fluctuations during
the inflationary scenario may play a fundamental  role in
producing the primordial fluctuations \cite{6252602,1126,6252603}. During the warm
inflationary scenario the density perturbations arise from
thermal fluctuations of the inflaton  and dominate over the
quantum ones. In this form,
an essential  condition for warm inflation to occur is the
existence of a radiation component with temperature $T>H$, since the thermal and quantum
fluctuations are proportional to $T$ and $H$,
respectively\cite{warm1,warm2,6252602,1126,6252603,6252604,62526,Moss:2008yb,Ramos:2013nsa}. When the universe heats
up and becomes radiation dominated, inflation ends and the
universe smoothly enters in the radiation
Big-Bang phase\cite{warm1}. For a comprehensive review of warm
inflation, see Ref. \cite{Berera:2008ar}. In this direction, there are many phenomenological models of warm inflation, but more interesting are the first principles model of warm inflation in which the dissipative coefficient and effective potential are computed from quantum field theory. For instance in Ref. \cite{Berera:2008ar}, it was considered the following superpotential $W\left( \Phi\right)$
\begin{equation}
W\left( \Phi\right)=\frac{\lambda}{p+1}\frac{\Phi^{p+1}}{m_p^{p-2}},\label{W}
\end{equation}
which reduces to chaotic inflation models for $p>1$. In particular, in Ref.\cite{Bartrum:2013fia} the authors studied the quartic potential $V(\phi)=\lambda \phi^4$, which corresponds to a superpotential $W\left( \Phi\right)=\lambda Phi^3/3$ together with the dissipative coefficient corresponding to $a=3$, i.e., $\Gamma \propto T^3/\phi^2$, was confronted with current data available at that time. On the other hand, when $p=0$ and $\lambda<0$, we have supersymmetric hybrid inflation.

Given the attractiveness of the power-law plateau inflation models as a new class of candidates in describing inflation, 
the main goal of this work is study the consequences of considering this new family of single-field inflation models in the warm inflation scenario in order to avoid the reheating phase. We would like to
emphasize that our analysis is phenomenological in the sense that, in order to describe the dissipative effects during the inflationary expansion, we consider the generalized expression for the inflaton decay rate, given by Eq.(\ref{G}) and without considering a first principles construction for our model. However, in Ref.\cite{Dimopoulos:2016zhy}, the authors presented a toy-model in supergravity (SUGRA) which can produce the scalar potential of Plateau inflation (\ref{Vplateau}) for $n=2$ and $q=1$. Specifically, they considered only global supersymmetry (SUSY) and sub-Planckian fields with the superpotential
\begin{equation}
W(\Phi_1,\Phi_2)=\frac{S^2\left(\Phi_1^2-\Phi_1^2\right)}{2m},\label{SW}
\end{equation}
where $S$, $\Phi_1$, $\Phi_2$ are chiral superfields and $m$ is a large, but sub-Planckian scale. An interesting approach could be specify a decaying mechanism for the scalar component of the superfields $\Phi_1$, $\Phi_2$ in light fields which form the thermal bath and compute the corresponding dissipative coefficient $\Gamma$ starting from first principles. However, this further considerations go beyond the scope of our present work, but these may be regarded as basis of a future work. On the other hand, we will restrict ourselves only to study the strong dissipative regime, $R\gg 1$. For this dissipative regime, under the slow-roll approximation, we 
study the background as well as the perturbative dynamics. The free parameters characterizing our model are constrained by considering the essential condition for warm inflation, $T>H$, the condition for the model evolves according to strong  dissipative regime, and the 2015 Planck results through the $n_s-r$ plane. For completeness, we study the predictions of our model regarding the running of the scalar spectral index, through the $n_s-dn_s/d \ln k$ plane. 

This paper is organized as follows: In the next section, we present
the basic setup of warm inflation. In section \ref{WPI} we study the background and perturbative dynamics
when our model evolves according to strong regime. Specifically, we find explicit expressions for the most relevant inflationary observables as the scalar power spectrum, scalar spectral index, the running of the scalar spectral index, and tensor-to-scalar ratio.
In order to establish a direct comparison between the power-law plateau inflation in the cold and warm scenarios, we will restrict ourselves to the case $n=2$ and $q=1$. For this particular case we obtain the predictions in the $r-n_s$ and $n_s-dn_s/d\ln k$ plane. 

Finally, section \ref{conclu} summarizes our finding and exhibits our conclusions. We have chosen units such that $c=\hbar=1$.

\section{Basics of warm inflation scenario}\label{WI}

In this section, we introduce the basic setup of warm inflation 

\subsection{Background evolution}

We start by considering a spatially flat Friedmann-Robertson-Walker (FRW) universe
containing a self-interacting inflaton scalar field $\phi$ with energy density and pressure
given by $\rho_{\phi}=\dot{\phi}^2/2+V(\phi)$ and $P_{\phi}=\dot{\phi}^2/2-V(\phi)$, respectively,
and a radiation field with energy density $\rho_{\gamma}$. The corresponding Friedmann equations reads
\begin{equation}
H^2=\frac{1}{3M^2_p}(\rho_{\phi}+\rho_{\gamma}),\label{Freq}
\end{equation}
where $M_p=\frac{1}{\sqrt{8\pi G}}$ is the reduced Planck mass.

The dynamics of $\rho_{\phi}$ and $\rho_{\gamma}$ is described by the equations \cite{warm1,warm2}
\begin{equation}
\dot{\rho_{\phi}}+3\,H\,(\rho_{\phi}+P_{\phi})=-\Gamma \dot{\phi}^{2},
\label{key_01}%
\end{equation}
and
\begin{equation}
\dot{\rho}_{\gamma}+4H\rho_{\gamma}=\Gamma \dot{\phi}^{2}, \label{key_02}%
\end{equation}
where the  dissipative coefficient $\Gamma>0$ produces the decay of the scalar
field into radiation. Recall that this
decay rate can be assumed  to be a function of the
temperature of the thermal bath $\Gamma(T)$, or a function of the
scalar field $\Gamma(\phi)$, or a function of $\Gamma(T,\phi)$ or
simply a constant. As we have mentioned in the introduction, the parametrization given by Eq.(\ref{G})
includes different cases, depending of the
values of $a$. Particularly, the inflaton decay rates $a=3$ ($\Gamma= C_{\phi}\frac{T^3}{\phi^{2}}$) and  $a=1$ ($\Gamma=C_{\phi}T$) have been studied extensively in the literature \cite{Panotopoulos:2015qwa,BasteroGil:2012cm,Benetti:2016jhf,yowarm1,yowarm2,yowarm3,yowarm4,yowarm5}.

During warm inflation, the energy density related to
the scalar field predominates  over the energy density of the
radiation field, i.e.,
$\rho_\phi\gg\rho_\gamma$\cite{warm1,warm2,6252602,1126,6252603,6252604,62526,Moss:2008yb}, but even if small when compared to the inflaton energy density
it can be larger than the expansion rate with $\rho_{\gamma}^{1/4}>H$. Assuming thermalization, this translates roughly
into $T>H$, which is the condition for warm inflation to occur.

When $H$, $\phi$, and $\Gamma$ are slowly varying, which is a good
approximation during inflation, the production of radiation becomes quasi-stable, i.e., $\dot{\rho
}_{\gamma}\ll4H\rho_{\gamma}$ and $\dot{\rho}_{\gamma}\ll\Gamma\dot{\phi}^{2}%
$, see Refs.\cite{warm1,warm2,6252602,1126,6252603,6252604,62526,Moss:2008yb}. Then, the equations of motion reduce to
\begin{equation}
3\,H\,(1+R)\dot{\phi}\simeq -V_{,\phi},
\label{key_01n}%
\end{equation}
where $,\phi$ denotes differentiation with respect to inflaton, and
\begin{equation}
4H\rho_{\gamma}\simeq \Gamma\,\dot{\phi}^{2}, \label{key_02n}%
\end{equation}
where $R$ is the dissipative ratio defined as
\begin{equation}
R\equiv\frac{\Gamma}{3H}.
\end{equation}

In warm inflation, we can distinguish between two possible scenarios, namely the weak and strong dissipative regimes, defined as $R\ll 1$ and $R\gg 1$, respectively. In the weak dissipative regime, the Hubble damping is still the dominant term, however, in the strong dissipative regime, the dissipative coefficient $\Gamma$ controls the damped evolution of the inflaton field.

If we consider thermalization, then the energy density of the radiation field could be written as $\rho_{\gamma}=C_{\gamma}\,T^{4}$, where the constant  $C_{\gamma}=\pi^{2}\,g_{\ast}/30$. Here,   $g_{\ast}$ represents the number
of relativistic degrees of freedom. In the Minimal Supersymmetric Standard Model (MSSM), $g∗ = 228.75$ and $C_{\gamma} \simeq 70$ \cite{62526}. Combining Eqs.(\ref{key_01n}) and (\ref{key_02n}) with $\rho_{\gamma}\propto\,T^{4}$, the temperature of the
thermal bath becomes
\begin{equation}
T=\left[\frac{\Gamma\,V_{,\phi}^2}{36 C_{\gamma}H^3(1+R)^2}\right]^{1/4}.\label{temp}
\end{equation}

 On the other hand, the consistency conditions for the approximations to hold imply that a set of slow-roll conditions must be satisfied for a prolonged period of inflation to take place. For warm inflation, the slow-roll parameters are \cite{26,62526}
\begin{equation}
\epsilon=\frac{M^2_p}{2}\left(\frac{V_{,\phi}}{V}\right)^2,\,\,\,\eta=M^2_p\left(\frac{V_{,\phi \phi}}{V}\right),\,\,\,\beta=M^2_p \left(\frac{\Gamma_{,\phi}\,V_{,\phi}}{\Gamma\,V}\right),\,\,\,\sigma = M_p^2 \left(\frac{V_{,\phi}}{\phi V}\right).\label{srparam}
\end{equation}

The slow-roll conditions for warm inflation can be expressed as \cite{26,62526,Moss:2008yb}
\begin{equation}
\epsilon \ll 1+R,\,\,\,\eta \ll 1+R,\,\,\,\beta \ll 1+R,\,\,\,\sigma\ll 1+R\label{srcon}
\end{equation}

When one these conditions is not longer satisfied, either the motion of the inflaton is no
longer overdamped and slow-roll ends, or the radiation becomes comparable to the inflaton energy density. In this way,
inflation ends when one of these parameters become the order of $1+R$.

The number of $e$-folds in the slow-roll approximation, using  (\ref{Freq}) and (\ref{key_01n}), yields
\begin{equation}
N \simeq -\frac{1}{M_p^2}\int_{\phi_{*}}^{\phi_{end}}\frac{V}{V_{,\phi}}(1+R)d\phi,\label{Nfolds}
\end{equation}
where $\phi_{*}$ and $\phi_{end}$ are the values of the scalar field when the cosmological scales crosses the Hubble-radius and at the end of inflation, respectively.
As it can be seen, the number of $e$-folds is increased due to an extra term
of $(1+R)$. This implies a more amount of inflation, between these two values of the field, compared to cold
inflation.

\subsection{Cosmological perturbations}

In the warm inflation scenario, a
thermalized radiation component is present with $T>H$, then the inflaton fluctuations
$\delta \phi$ are predominantly thermal instead quantum. In this way, following \cite{1126,62526,Moss:2008yb,Berera:2008ar}, the
amplitude of the power spectrum of the curvature perturbation is given by

\begin{equation}
{\cal{P}_{\cal{R}}}^{1/2}\simeq \left(\frac{H}{2\pi}\right) \left(\frac{3H^2}{V_{,\phi}}\right)\left(1+R\right)^{5/4}\left(\frac{T}{H}\right)^{1/2},\label{PR}
\end{equation}
where the normalization has been chosen in order to recover the standard cold inflation result when $R\rightarrow 0$ and $T \simeq H$.

By the other hand, the scalar spectral index $n_s$ to leading order in the slow-roll approximation, is given by \cite{62526,Moss:2008yb}
\begin{equation}
n_s=1+\frac{d\ln{\cal{P}_{\cal{R}}}}{d\ln k}\simeq 1-\frac{(17+9R)}{4(1+R)^2}\epsilon-\frac{(1+9R)}{4(1+R)^2}\beta+\frac{3}{2(1+R)}\eta.\label{nsw}
\end{equation}

We also introduce the running of the scalar spectral index, which represents the scale
dependence of the spectral index, by $n_{run}=\frac{d n_s}{d \ln k}$. In particular, for the strong dissipative regime, this expressions is given by \cite{62526,Moss:2008yb}
\begin{equation}
\frac{d n_s}{d \ln k}\simeq \frac{1}{R^2}\left(-\frac{9}{2}\beta^2-\frac{27}{4}\epsilon^2-\frac{9}{2}\epsilon \beta
+\frac{15}{4}\eta \beta +6 \epsilon \eta -\frac{3}{2}\zeta^2+\frac{9}{2}\gamma \epsilon\right),\label{nrun}
\end{equation}
where $\zeta^2$ and $\gamma$ are second-order slow-roll parameters defined by
\begin{equation}
\zeta^2\equiv M_p^4\left(\frac{V_{,\phi} V_{,\phi \phi \phi}}{V^2}\right),\label{zeta}
\end{equation}
and 
\begin{equation}
\gamma \equiv M_p^2\left(\frac{\Gamma_{,\phi \phi}}{\Gamma}\right),\label{gam}
\end{equation}
respectively.

Regarding to tensor perturbations, these do not couple to the thermal background, so gravitational waves are only generated by quantum fluctuations, as
in standard inflation \cite{Ramos:2013nsa}. However, the tensor-to-scalar ratio $r$ is modified with respect to standard cold inflation, yielding \cite{Berera:2008ar}
\begin{equation}
r\simeq \left(\frac{H}{T}\right)\frac{16\epsilon}{(1+R)^{5/2}}.\label{rwi}
\end{equation}
We can see that warm inflation predicts a tensor-to-scalar ratio suppressed by a factor $(T/H)(1 + R)^{
5/2} > 1$ compared with standard cold inflation.

When a specific form of the scalar potential and the dissipative coefficient $\Gamma$ are considered, it is possible to study the
background evolution under the slow-roll regime and the primordial perturbations in order to
test the viability of warm inflation. In the following we will study how an inflaton decay rate with a generic power-law dependence with the scalar field $\phi$ and the temperature of the thermal bath $T$ influences the inflationary dynamics for the power-law plateau potential. We will restrict ourselves to the strong dissipation regime.

\section{Dynamics of warm power-law plateau inflation in the strong dissipative regime}\label{WPI}

\subsection{Background evolution}

Assuming that the inflationary dynamics takes place in the strong dissipative regime, i.e., $R\gg1$ (or
$\Gamma \gg 3H$), by using Eqs. (\ref{G}) and (\ref{temp}), the temperature of the thermal bath
as function of the inflaton field is found to be
\begin{equation}
T= \left[4\sqrt{3}C_{\phi}^{-1}\left(nq\alpha^n\right)^2\right]^{\frac{1}{4+a}}\left(V_0^{\frac{3}{2}}M_p^{a-2}\right)^{\frac{1}{4+a}}
\left(\frac{\phi}{M_p}\right)^{\frac{3(nq-2)+2a}{2(4+a)}}\left(\alpha^n+\left(\frac{\phi}{M_p}\right)^n\right)^{-\frac{(4+3q)}{2(4+a)}}.\label{Tphi}
\end{equation}
Replacing the last expression into Eq.(\ref{G}), both the inflaton decay rate and the ratio $R=\Gamma/3H$ expressed in terms of the inflaton field becomes
\begin{equation}
\Gamma= \left[\left(\sqrt{3}(4C_{\gamma})^{-1}\left(nq\alpha^n\right)^2\right)^a C_{\phi}^4\right]^{\frac{1}{4+a}}\left(V_0^{\frac{3a}{2}}M_p^{4-5a}\right)^{\frac{1}{4+a}}
\left(\frac{\phi}{M_p}\right)^{\frac{8+3a(nq-4)}{2(4+a)}}\left(\alpha^n+\left(\frac{\phi}{M_p}\right)^n\right)^{-\frac{a(4+3q)}{2(4+a)}},\label{Gammaphi}
\end{equation}
and
\begin{equation}
R= \left[\left(\sqrt{3}(4C_{\gamma})^{-1}\left(nq\alpha^n\right)^2\right)^a C_{\phi}^4\right]^{\frac{1}{4+a}}\left(\frac{M_p^4}{V_0}\right)^{\frac{2-a}{4+a}}
\left(\frac{\phi}{M_p}\right)^{\frac{8+3a(nq-4)}{2(4+a)}-\frac{nq}{2}}\left(\alpha^n+\left(\frac{\phi}{M_p}\right)^n\right)^{-\frac{a(4+3q)}{2(4+a)}+\frac{nq}{2}},\label{Rphi}
\end{equation}
respectively.

In this way, by combining Eqs.(\ref{Vplateau}), (\ref{key_01n}), and (\ref{Gammaphi}), the inflaton field as function of cosmic time may be obtained from the following expression
\begin{eqnarray}
\left[\sqrt{3}(4C_{\gamma})^{-a}\left(n\alpha^n\right)^{-8}C_{\phi}^4\,q^{a-4}\right]^{\frac{1}{4+a}}g_{0}^{-1}\left(V_0^{-\frac{1}{2}-\frac{6}{4+a}}\,M_p^{\frac{3(4-a)}{4+a}}\right)\left(\frac{\phi}{M_p}\right)^{ng_0}\\\nonumber
 \times \left(\alpha^n+\left(\frac{\phi}{M_p}\right)^n\right)^{\frac{q(8-a)+16}{2(4+a)}}\,_{2}F_{1}\left[1,\frac{12-4a+8n}{n(4+a)},1+g_0,-\alpha^{-n}(\phi/M_p)^n\right] &=&-t,\label{phit}
\end{eqnarray}
where $g_0\equiv \frac{24-8nq-8a+anq}{2n(4+a)}$ and $_2F_1$ denotes the hypergeometric function \cite{arfken}.

For this model the set of slow-roll parameters become
\begin{eqnarray}
\epsilon&=&\frac{\left(nq\alpha^n\right)^2}{2\left(\frac{\phi}{M_p}\right)^2\left(\alpha^n+\left(\frac{\phi}{M_p}\right)^n\right)^2},\label{eps}\\
\eta&=&\frac{nq\alpha^n}{\left(\frac{\phi}{M_p}\right)^2}\frac{\left[(nq-1)\alpha^n-(n+1)\left(\frac{\phi}{M_p}\right)^n\right]}{\left(\alpha^n+\left(\frac{\phi}{M_p}\right)^n\right)^2},\label{et}\\
\beta &=&\frac{nq\alpha^n}{2(4+a)\left(\frac{\phi}{M_p}\right)^2}\frac{\left[(8+3a(nq-4))\alpha^n-4(a(3+n)-2)\left(\frac{\phi}{M_p}\right)^n\right]}{\left(\alpha^n+\left(\frac{\phi}{M_p}\right)^n\right)^2},\label{bet}\\
\sigma&=&\frac{nq\alpha^n}{\left(\frac{\phi}{M_p}\right)^2\left(\alpha^n+\left(\frac{\phi}{M_p}\right)^n\right)}.\label{sig}
\end{eqnarray}

For the strong dissipative regime, the slow-roll conditions (\ref{srcon}) become
\begin{equation}
\epsilon \ll R,\,\,\,\eta \ll R,\,\,\,\beta \ll R,\,\,\,\sigma\ll R.\label{srcons}
\end{equation}
As we mentioned in the previous section, inflation ends when one of these parameters become the order of $R$.

On the other hand, the number of inflationary $e$-folds between the values of the scalar field when a given
perturbation scale leaves the Hubble-radius and at the end of inflation, can be computed from Eqs.(\ref{Vplateau}) and (\ref{Rphi}) into (\ref{Nfolds}), yielding
\begin{eqnarray}
N&=&g_{1}^{-1}\left[\left(9(4C_{\gamma})^a\left(n\alpha^n\right)^8\right)^{-1}\,C_{\phi}^4\,q^{a-4}\right]^{\frac{1}{4+a}}\left(\frac{M_p^4}{V_0}\right)^{\frac{2-a}{4+a}}\left(\frac{\phi}{M_p}\right)^{ng_1}\nonumber\\
&&\times \left(\alpha^n+\left(\frac{\phi}{M_p}\right)^n\right)^{\frac{8+q(2-a)}{4+a}}\,_2F_1\left[1,\frac{12-4a+8n}{n(4+a)},1+g_1,-\alpha^{-n}(\phi/M_p)^n\right] \bigg |_{\phi_{end}}^{\phi_{*}},\label{Nphi}
\end{eqnarray}
where $g_1\equiv \frac{nq(a-2)-4(a-3)}{n(4+a)}$. 

Since Eq.(\ref{Nphi}) has a complicated dependence in the inflaton field, it is not possible to express 
$\phi_{*}$ as function of $N$ analytically. Instead, for numerical purposes, from Eq.(\ref{Vplateau}) we may express the inflaton field as function of the amplitude of the potential and evaluate this expression at the Hubble-radius crossing, obtaining
\begin{equation}
\label{phistar}
\frac{\phi_{*}}{M_p}=\left[\frac{\alpha^n \left(\frac{V_{*}}{V_0}\right)^{\frac{1}{q}}}{1-\left(\frac{V_{*}}{V_0}\right)^{\frac{1}{q}}}\right]^{\frac{1}{n}}.
\end{equation}

Last expression will be useful to evaluate the several inflationary observables and put the observational bounds on our model.

\subsection{Cosmological perturbations}

Now, we shall study the cosmological perturbations for our model in the strong dissipative regime
$R=\Gamma/3H>1$. For this regime, the amplitude of the scalar power spectrum (\ref{PR}) becomes
\begin{equation}
{\cal{P}_{\cal{R}}}^{1/2}\simeq \left(\frac{H}{2\pi}\right) \left(\frac{3H^2}{V_{,\phi}}\right)\left(\frac{T}{H}\right)^{1/2}R^{5/4},\label{PR}
\end{equation}
then, by replacing Eqs.(\ref{Vplateau}), (\ref{Tphi}), and (\ref{Rphi}), the power spectrum as function of the inflaton field is found to be
\begin{eqnarray}
{\cal{P}_{\cal{R}}}=\frac{\left(2^7C_{\gamma}^{5/2}\right)^{-1}}{\pi^2}\left[2^{18}\sqrt{3}^{-(13+a)}\left(C_{\gamma}C_{\phi}\right)^9\,\left(nq\alpha^n\right)^{3(a-2)}\right]^{\frac{1}{4+a}}\left(\frac{M_p^4}{V_0}\right)^{\frac{3(1-2a)}{2(4+a)}}\nonumber\\
 \times \left(\frac{\phi}{M_p}\right)^{\frac{6(5-a)-3nq(1-2a)}{2(4+a)}}\left(\alpha^n+\left(\frac{\phi}{M_p}\right)^n\right)^{\frac{3\left[2(2-a)+q(1-2a)\right]}{2(4+a)}}.
\end{eqnarray}

By considering the strong dissipative regime, the expression for the scalar spectral index becomes
\begin{equation}
n_s\simeq 1+\frac{1}{R}\left(-9\epsilon-9\beta +6\eta\right),
\end{equation}
in this way, by replacing Eqs.(\ref{Rphi}), and (\ref{eps})-(\ref{bet}), the inflaton field depende of the scalar spectral index is given by
\begin{eqnarray}
\label{nsf}
n_s &=& 1+\frac{1}{4+a}\left[2^{a-4}3^{6+a}C_{\gamma}^aC_{\phi}^{-4}\left(nq\alpha^n\right)^{4-a}\right]^{\frac{1}{4+a}}\left(\frac{M_p^4}{V_0}\right)^{\frac{a-2}{4+a}}\left(\frac{\phi}{M_p}\right)^{\frac{4(a-3)-nq(a-2)}{4+a}} \left(\alpha^n+\left(\frac{\phi}{M_p}\right)^n\right)^{\frac{-8+q(a-2)}{4+a}}\nonumber\\
&&\times \left[(2(4a-5)+nq(1-2a))\alpha^n+2(4a-5+n(a-2)) \left(\frac{\phi}{M_p}\right)^n \right].
\end{eqnarray}

Regarding the running of the scalar spectral index, the second-order slow-roll parameters $\zeta^2$ and $\gamma$ for this model become
\begin{eqnarray}
\zeta^2 &=& \frac{\left(nq\alpha^n\right)^2}{\left(\frac{\phi}{M_p}\right)^4\left(\alpha^n+\left(\frac{\phi}{M_p}\right)^n\right)^4}
\bigg[\left(2-3nq+n^2q^2\right)\alpha^{2n}-(1+n)(n-4+3nq)\alpha^n\left(\frac{\phi}{M_p}\right)\nonumber \\
&&+\left(2+3n+n^2\right)\left(\frac{\phi}{M_p}\right)^{2n}\bigg],\label{zetaphi}
\end{eqnarray}
and
\begin{eqnarray}
\gamma &=& \frac{a}{4(4+a)^2\left(\frac{\phi}{M_p}\right)^2\left(\alpha^n+\left(\frac{\phi}{M_p}\right)^n\right)^2}\bigg[(3nq-14)(8+3a(nq-4))\alpha^{2n}\nonumber\\
&& -2\left(4\left(28+n(4-3q)+n^2(4+3q)\right)+a\left(-168+13n(3q-4)+n^2(4+15q)\right)\right)\alpha^n\left(\frac{\phi}{M_p}\right)^{n}\nonumber\\
&& +8(7+2n)(a(n+3)-2)\left(\frac{\phi}{M_p}\right)^{2n}\bigg],\label{gaphi}
\end{eqnarray}
respectively. Then, by replacing last expressions together with Eqs.(\ref{eps})-(\ref{bet}) into (\ref{nrun}), the running of the spectral index is completely determined (not shown).

Regarding the tensor perturbations, the tensor-to-scalar ratio for the strong regime becomes
\begin{equation}
r\simeq \left(\frac{H}{T}\right)\frac{16\epsilon}{R^{5/2}}.\label{rs}
\end{equation}
The inflaton field dependence of the tensor-to-scalar ratio is determined by replacing Eqs.(\ref{Vplateau}), (\ref{temp}), and (\ref{eps})
into (\ref{rs}), yielding 
\begin{eqnarray}
r &=& 2^8C_{\gamma}^{5/2}\left[2^{-18}\sqrt{3}^{5-a}\left(C_{\gamma}\,C_{\phi}\right)^{-9}\left(nq\alpha^n\right)^{3n(2-a)}\right]^{\frac{1}{4+a}}\left(\frac{M_p^4}{V_0}\right)^{\frac{4a-11}{2(4+a)}}\nonumber\\
&& \times \left(\frac{\phi}{M_p}\right)^{\frac{-30+11nq-4a(nq-6)}{2(4+a)}}\left(\alpha^n+\left(\frac{\phi}{M_p}\right)^n\right)^{\frac{6(a-2)+q(4a-11)}{2(4+a)}}.\label{rrphi}
\end{eqnarray}

In order to find observational constraints on our model, we will study the particular case $n=2$ and $q=1$, which corresponds to the best choice of model in the power-law plateau inflation in Ref.\cite{Dimopoulos:2016zhy}. In addition, we consider $\alpha$ and $V_0$, from the potential (\ref{Vplateau}), and $C_{\phi}$, from the generalized inflaton decay ratio (\ref{G}), as free parameters characterizing the model.

\subsection{Special case $n=2$ and $q=1$}

To compare the predictions of power-law plateau inflation in the cold and warm scenarios, we will restrict ourselves to the case 
$n=2$ and $q=1$, corresponding to the best choice of model in Ref.\cite{Dimopoulos:2016zhy}. Moreover, for the generic parametrization of the inflaton decay rate (\ref{G}), we consider the cases $a=3, 1, 0 ,$ and $-1$, which correspond to several dissipative ratios studied in the literature, but we consider $\alpha$, $V_0$ and $C_{\phi}$ to be free parameters. To put observational constraints on the parameters characterizing our model, we consider the essential condition for warm inflation, $T>H$, the condition for which the model evolves according to the strong regime, $R\gg 1$, and finally the two-dimensional marginalized joint confidence contours for $n_s$ and $r$, at the 68 and 95 $\%$ CL and the amplitude of the scalar power spectrum by Planck 2015 data \cite{Ade:2015lrj}. In addition,  we will try to
ascertain whether the predictions for the running of the scalar spectral index $d n_s/d \ln k$ are consistent with
the current bounds imposed by Planck.

\subsubsection{$a=3$}

In first place, for the special case $a=3$, i.e., for $\Gamma\propto T^3/\phi^2$, the scalar spectral index (\ref{nsf}) becomes
\begin{equation}
n_s=1+\frac{\left(3^9C_{\gamma}^3 C_{\phi}^{-4}\right)^{\frac{1}{7}}}{7}\left(\frac{M_p^4}{V_0}\right)^{\frac{2}{7}}\left(\frac{\phi}{M_p}\right)^{-\frac{2}{7}}\frac{\left(4\alpha^2+18\left(\frac{\phi}{M_p}\right)^2\right)}{\left(\alpha^2+\left(\frac{\phi}{M_p}\right)^2\right)}.
\end{equation}
From last expression we see that $n_s$ is always greater that one. Based on current observational data, for $\Lambda$CDM
$+r+dn_s/d\ln k$, the spectral index is measured to be $n_s=0\textup{.}9667 \pm $ (68 \% CL, \emph{Planck} TT + LowP). Hence,
the inflaton decay ratio  corresponding to $a=3$ is not suitable to describe a strong dissipative warm inflation scenario consistent with current observations. It is interesting to mention that, for other inflaton potetentials, the inflaton decay ratio $a=3$ describes a consistent warm inflationary dynamics (see Refs.\cite{BasteroGil:2012cm,Benetti:2016jhf,yowarm1,yowarm2,yowarm3,yowarm4,yowarm5}).

\subsubsection{$a \neq 3$}

\begin{figure}[th]
{{\hspace{0cm}\includegraphics[width=3.21 in,angle=0,clip=true]{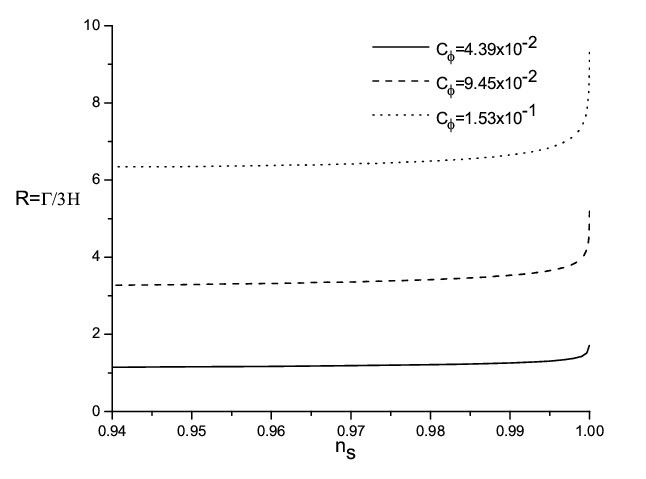}}}
{\includegraphics[width=3.21 in,angle=0,clip=true]{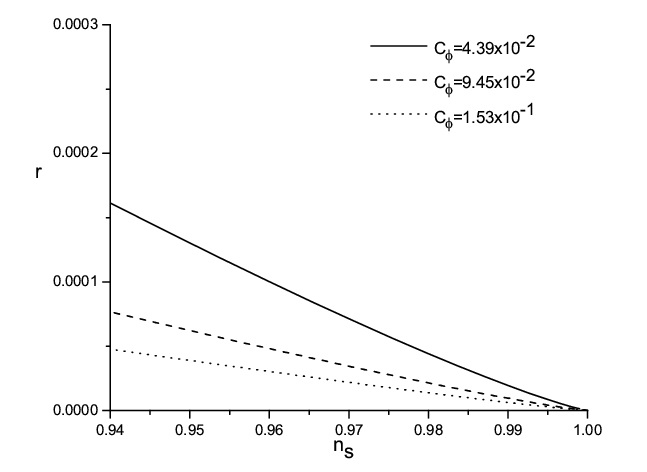}}

{\vspace{-0.5 cm}\caption{ Plots of $R=\Gamma/3H$ as function of
the scalar spectral index $n_s$ (left) and the tensor-to-scalar $r$  as function of
the scalar spectral index $n_s$ (right). For both plots we have
considered three different values of the parameter $C_\phi$ for the
case $a=1$, i.e., $\Gamma\propto T$, assuming 
the model evolves according to the strong dissipative regime. In both panels, the dotted, dashed, and
solid lines correspond to the pairs of values ($\alpha=0\textup{.}25$, $V_0=1\textup{.}03 \times 10^{-12}$),
($\alpha=0\textup{.}45$, $V_0=1\textup{.}09 \times 10^{-12}$), and ($\alpha=0\textup{.}65$, $V_0=1\textup{.}19 \times 10^{-12}$),
respectively. In these plots we have fixed the values
$C_\gamma=70$ and $M_p=1$.
 \label{fig1}}}
\end{figure}

Fig.\ref{fig1} shows the ratio $\Gamma/3H$ and the tensor-to-scalar ratio $r$ as functions of the scalar spectral index $n_s$ for the case $a=1$, i.e., $\Gamma (\phi,T)=C_{\phi}T$. To obtain the values to perform the plots we have used three different values for $C_{\phi}$ parameter and fixed the values $C_{\gamma}=70$ and $M_p=1$. For each value of $C_{\phi}$ we solve numerically the Eqs.(\ref{PR}) and (\ref{nsf}) (after evaluating both equations at $\phi_{*}$ given by Eq.(\ref{phistar}), which is a function of $V_{*}$) for $\alpha$ and $V_0$, considering the observational values
$\mathcal{P}_{\mathcal{R}}\simeq 2\times 10^{-9}$ and $n_s\simeq 0\textup{.}9667$ \cite{Ade:2015lrj}, and fixing $V_{*}=10^{-12}$. In this way, for $C_{\phi}=4\textup{.}39\times 10^{-2}$, we obtain the values $\alpha=0\textup{.}25$ and $V_0=1\textup{.}03 \times 10^{-12}$, whereas for $C_{\phi}=9\textup{.}45\times 10^{-2}$, the solution is given by $\alpha=0\textup{.}45$ and $V_0=1\textup{.}09 \times 10^{-12}$. Finally, for $C_{\phi}=1\textup{.}53\times 10^{-1}$, we found that $\alpha=0\textup{.}65$ and $V_0=1\textup{.}19 \times 10^{-12}$. In this way, the $R(n_s)$ and $r(n_s)$ curves
of Fig.(\ref{fig1}) may be generated by plotting Eqs.(\ref{nsf}), (\ref{Rphi}), and (\ref{rrphi}) parametrically (after being evaluated at $\phi_{*}$ given by Eq.(\ref{phistar})) with respect to $V_{*}$. 

\begin{figure}[th]
{{\hspace{0cm}\includegraphics[width=3.5 in,angle=0,clip=true]{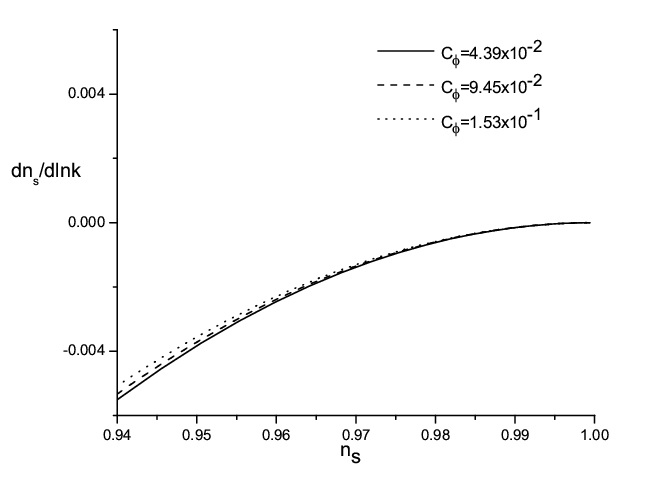}}}

{\vspace{-0.5 cm}\caption{ Plots of the running of the scalar spectral index $d n_s/d\ln k$ as function of
the scalar spectral index $n_s$. For this plot we have
considered three different values of the parameter $C_\phi$ for the
case $a=1$, i.e., $\Gamma\propto T$, assuming 
the model evolves according to the strong dissipative regime. The dotted, dashed, and
solid lines correspond to the pairs of values ($\alpha=0\textup{.}25$, $V_0=1\textup{.}03 \times 10^{-12}$),
($\alpha=0\textup{.}45$, $V_0=1\textup{.}09 \times 10^{-12}$), and ($\alpha=0\textup{.}65$, $V_0=1\textup{.}19 \times 10^{-12}$),
respectively. Additionally, we have fixed the values
$C_\gamma=70$ and $M_p=1$.
 \label{fig2}}}
\end{figure}

From the left panel, we observe that for $C_{\phi}\geq 4\textup{.}39\times 10^{-2}$, the model evolves according to the strong regime, $R>1$. On the other hand, we find numerically that, for $C_{\phi}\geq 4\textup{.}39\times 10^{-2}$, the ratio $\frac{T}{H}$ becomes $\frac{T}{H} \gtrsim 79$ when $n_s\simeq 0\textup{.}9667$ (plot not shown). Hence, for $C_{\phi}\geq 4\textup{.}39\times 10^{-2}$ the essential condition for warm inflation, $\frac{T}{H}>1$, is always satisfied. Then, the condition for which the model evolves in agreement with the strong regime gives us an lower limit on $C_{\phi}$. However, the essential condition for warm inflation does not impose any constraint on $C_{\phi}$.  On the other hand, right panel of Fig.\ref{fig1} shows the trajectories in the $n_s-r$ plane along with the two-dimensional marginalized constraints at 68 $\%$ and 95 $\%$ C.L. on the parameters $r$ and $n_s$, by Planck 2015 data  \cite{Ade:2015lrj}. Here, we observe that for $C_{\phi}\geq 4\textup{.}39\times 10^{-2}$, the tensor-to-scalar ratio predicted by
this model ratio is always consistent with the observational bound found by Planck, given by $r<0\textup{.}168$ ($95\%$ CL,  \emph{Planck} TT + LowP). In order to determine the prediction of this model regarding the running of the spectral
index, Fig.\ref{fig2} shows the trajectories in the $n_s-d n_s/d\ln k$ plane. Again we note that for $C_{\phi}\gtrsim 4\textup{.}39\times 10^{-2}$ the running of the spectral index predicted by the model is consistent with the bound found by Planck, given by $dn_s/d\ln k=-0\textup{.}0126^{+0\textup{.}0098}_{-0\textup{.}0087}$ ($68\%$ CL,  \emph{Planck} TT + LowP). After the previous analysis, we only were able to find a lower limit on $C_{\phi}$ as well as for $\alpha$ and $V_{0}$, given by given by $C_{\phi}=4\textup{.}39\times 10^{-2}$, $\alpha=0\textup{.}25$ and $V_0=1\textup{.}03 \times 10^{-12}$.

\begin{figure}[th]
{{\hspace{0cm}\includegraphics[width=3.21 in,angle=0,clip=true]{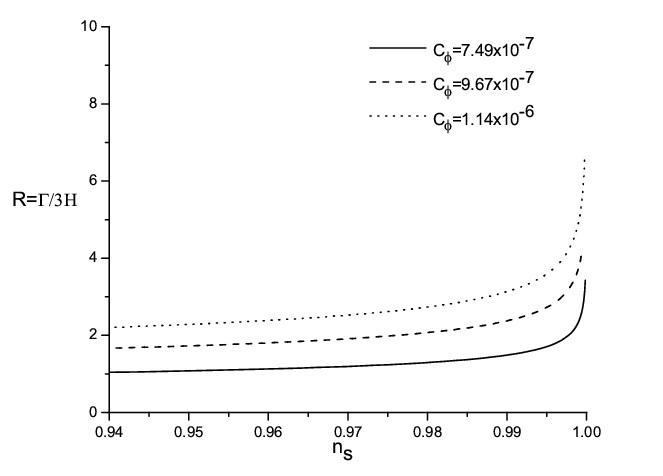}}}
{\includegraphics[width=3.21 in,angle=0,clip=true]{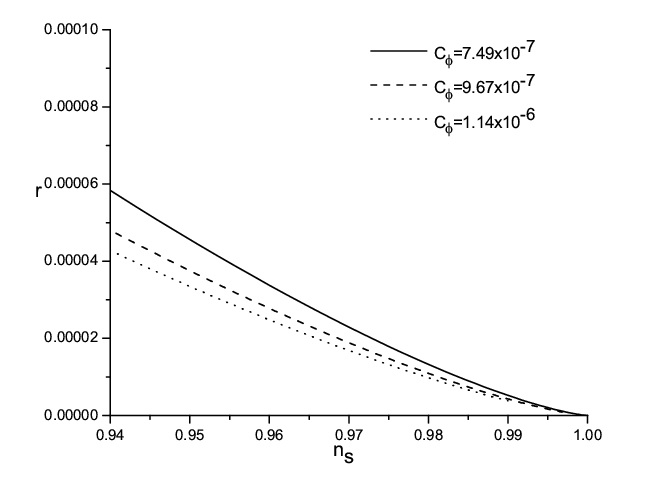}}

{\vspace{-0.5 cm}\caption{ Plots of $R=\Gamma/3H$ as function of
the scalar spectral index $n_s$ (left) and the tensor-to-scalar $r$  as function of
the scalar spectral index $n_s$ (right). For both plots we have
considered three different values of the parameter $C_\phi$ for the
case $a=0$, i.e., $\Gamma\propto \phi$, assuming 
the model evolves according to the strong dissipative regime. In both panels, the dotted, dashed, and
solid lines correspond to the pairs of values ($\alpha=0\textup{.}4$, $V_0=1\textup{.}02 \times 10^{-12}$),
($\alpha=0\textup{.}8$ and $V_0=1\textup{.}06 \times 10^{-12}$), and ($\alpha=1\textup{.}2$, $V_0=1\textup{.}11 \times 10^{-12}$),
respectively. In these plots we have fixed the values
$C_\gamma=70$ and $M_p=1$.
 \label{fig3}}}
\end{figure}

The same analysis can be done for the case $a=0$, for which $\Gamma \propto \phi$. Fig.\ref{fig3} shows the ratio $\Gamma/3H$ and the tensor-to-scalar ratio $r$ as functions of the scalar spectral index. To obtain the values to perform the plots, we have used 
three different values for $C_{\phi}$ and followed the same procedure as the case $a=1$, solving numerically Eqs.(\ref{PR}) and (\ref{nsf}) (after evaluating both equations at $\phi_{*}$ given by Eq.(\ref{phistar})) for $\alpha$ and $V_0$, considering $\mathcal{P}_{\mathcal{R}}\simeq 2\times 10^{-9}$, $n_s\simeq 0\textup{.}9667$ \cite{Ade:2015lrj}, and fixing $V_{*}=10^{-12}$. In this way, for $C_{\phi}=7\textup{.}49\times 10^{-7}$, we obtain the values $\alpha=0\textup{.}4$ and $V_0=1\textup{.}02 \times 10^{-12}$, whereas for $C_{\phi}=9\textup{.}67\times 10^{-7}$, the solution is given by $\alpha=0\textup{.}8$ and $V_0=1\textup{.}06 \times 10^{-12}$. Finally, for $C_{\phi}=1\textup{.}14\times 10^{-6}$, we found that $\alpha=1\textup{.}2$ and $V_0=1\textup{.}11 \times 10^{-12}$. In this way, the $R(n_s)$ and $r(n_s)$ curves
of Fig.(\ref{fig1}) may be generated by plotting Eqs.(\ref{nsf}), (\ref{Rphi}), and (\ref{rrphi}) parametrically with respect to $V_{*}$. 

 From left panel of Fig.(\ref{fig3}), the condition for the model evolves according to strong regime is satisfied for $C_{\phi} \geq 7\textup{.}49\times 10^{-7}$, which gives us a lower limit for $C_{\phi}$. Additionally, for $C_{\phi} \geq 7\textup{.}49\times 10^{-7}$ the condition for warm inflation, $\frac{T}{H}>1$, is always satisfied. In particular, for $C_{\phi} =7\textup{.}49\times 10^{-7}$, the ratio $\frac{T}{H}$ becomes $\frac{T}{H} \simeq 55$ when $n_s\simeq 0\textup{.}9667$. Then, just like the case $a=1$, for $a=0$ the essential condition for warm inflation does not impose any constraint on $C_{\phi}$. Moreover, from the right panel, for $C_{\phi} \geq 7\textup{.}49\times 10^{-7}$, the tensor-to-scalar ratio becomes $r\sim 10^{-5}$, but this value is still supported by the last data of Planck. For completeness, the running of the scalar spectral index becomes $dn_s/d\ln k \simeq -0\textup{.}001$ at $n_s\simeq 0\textup{.}9667 $ (plot not shown). Then, for the case $a=0$, the previous analysis gives us only a lower limit for $C_{\phi}$ as well as for $\alpha$ and $V_0$, given by $C_{\phi}=7\textup{.}49\times 10^{-7}$, $\alpha=0\textup{.}4$, and $V_0=1\textup{.}02 \times 10^{-12}$ respectively. Despite this result, it is interesting to mention that for this power-law plateau potential, the inflaton decay ratio $\Gamma \propto \phi$ describes a strong dissipative warm inflation scenario compatible with current observations. In previous works
\cite{yowarm1,yowarm2,yowarm3,yowarm4,yowarm5} it was found that the decay rate $\Gamma \propto \phi$ is not able to describe a consistent strong dissipate dynamics, since the predicted scalar spectral index is always greater than unity.
 
 \begin{figure}[th]
{{\hspace{0cm}\includegraphics[width=3.21 in,angle=0,clip=true]{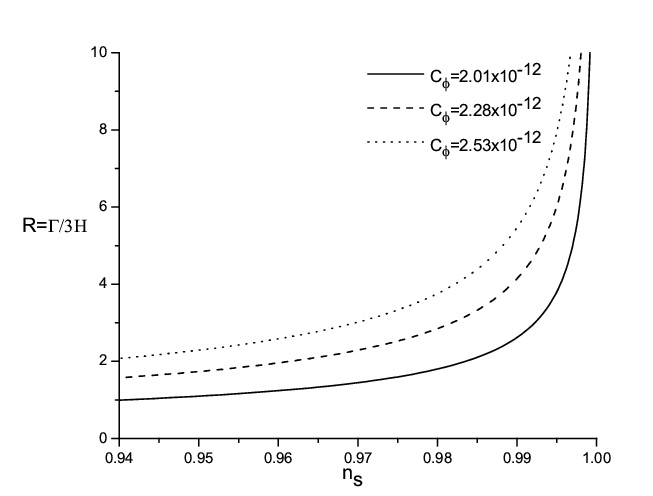}}}
{\includegraphics[width=3.21 in,angle=0,clip=true]{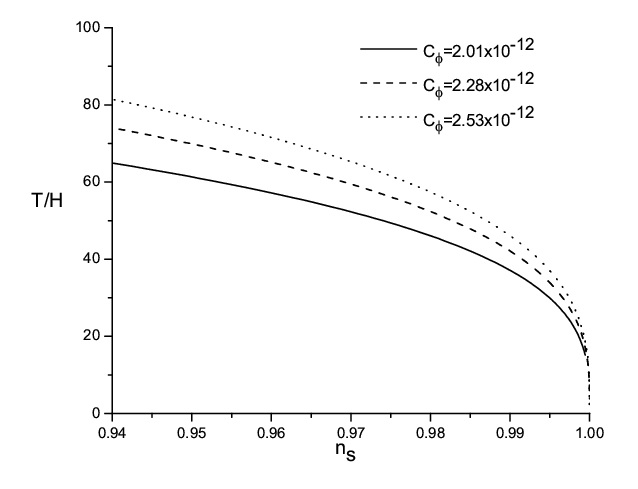}}

{\vspace{-0.5 cm}\caption{ Plots of the ratios $R=\Gamma/3H$ (left) and
$T/H$ (right), both as functions of
the scalar spectral index $n_s$ (right). For both plots we have
considered three different values of the parameter $C_\phi$ for the
case $a=-1$, i.e., $\Gamma\propto \phi^2/T$, assuming 
the model evolves according to the strong dissipative regime. In both panels, the dotted, dashed, and
solid lines correspond to the pairs of values ($\alpha=1\textup{.}5$, $V_0=1\textup{.}07 \times 10^{-12}$),
($\alpha=1\textup{.}95$ and $V_0=1\textup{.}09 \times 10^{-12}$), and ($\alpha=2\textup{.}4$, $V_0=1\textup{.}13 \times 10^{-12}$),
respectively. In these plots we have fixed the values
$C_\gamma=70$ and $M_p=1$.
 \label{fig4}}}
\end{figure}
 
Following the same procedure as the previous cases, for $a=-1$ we considered three different values for $C_{\phi}$. For $C_{\phi}=2\textup{.}01\times 10^{-12}$, we obtain the values $\alpha=1\textup{.}5$ and $V_0=1\textup{.}07 \times 10^{-12}$, whereas for $C_{\phi}=2\textup{.}28\times 10^{-12}$, the solution is given by $\alpha=1\textup{.}95$ and $V_0=1\textup{.}09 \times 10^{-12}$. Finally, for $C_{\phi}=2\textup{.}53\times 10^{-12}$, we found that $\alpha=2\textup{.}4$ and $V_0=1\textup{.}13 \times 10^{-12}$. Fig.(\ref{fig4}) shows the plots of the ratios $R=\Gamma/3H$ and $T/H$ as functions of the scalar spectral index $n_s$.
From left panel, we see that for $C_{\phi}\geq 2\textup{.}01\times 10^{-12}$ the model takes place in the strong dissipative regime
of warm inflation. Moreover, from right panel, we observe that the essential condition for warm inflation $\frac{T}{H}>1$ is always guaranteed. In particular, for $C_{\phi}=2\textup{.}01\times 10^{-12}$, this ratio takes the value $\frac{T}{H}\simeq 60$ when $n_s\simeq 0\textup{.}9667$. Again, the $\frac{T}{H}$ plot does not impose any constraint on $C_{\phi}$. Regarding the predictions of this case in the $r-n_s$ plane, for $C_{\phi}=2\textup{.}01\times 10^{-12}$, the tensor-to-scalar ratio becomes $r\sim 10^{-5}$, but this value is still supported by the last data of Planck by considering the two-dimensional marginalized joint confidence contours for $(n_s,r)$, at the 68 and 95 $\%$ C.L. (plot not shown). Finally, the predictions for the running of the spectral index are similiar to previous ones, yielding $d n_s/d \ln k\simeq -0\textup{.}001$ at $n_s\simeq 0\textup{.}9667$ for all the values considered for $C_{\phi}$ (plot not shown). The result of this analysis yields only a lower limit for $C_{\phi}$ as well as for $\alpha$ and $V_0$, given by $C_{\phi}=2\textup{.}01\times 10^{-12}$, $\alpha=1\textup{.}5$, and $V_0=1\textup{.}07 \times 10^{-12}$ respectively. Just like the case $a=0$, the case $a=-1$ has an interesting feature, because yields a strong dissipative dynamics compatible with observations, 
since that in previous works \cite{28,PRD,yowarm1,yowarm2,yowarm3,yowarm4,yowarm5}, the inflaton decay rate $\Gamma \propto \frac{\phi^2}{T}$ is not able to describe a consistent strong dissipative dynamics.

\subsection{Discussion}

From the analysis carried out in Ref.\cite{Dimopoulos:2016zhy}, and provided that the model be distinguishable from monomial inflation, i.e., $\phi >\alpha M_p$, the authors found that the best choice of model in the power-law plateau inflation family correspond to the values $n=2$ and $q=1$. In a first approach and ensuring a sub-Planckian excursion for the inflaton through the potential, the maximum value allowed for $\alpha$ was found to be $\alpha=0\textup{.}04$, and the values for the scalar spectral index and tensor-to-scalar ratio at the Hubble-radius crossing correspond to $n_s=0\textup{.}97$ and $r=0\textup{.}000157$. Four our warm power-plateau model, in order to produce
a strong dissipative dynamics, all the values obtained for $\alpha$ , for each value of $a$, are greater than $0\textup{.}04$, implying a trans-Planckian excursion of the inflaton field, but ensuring that $\phi >\alpha M_p$. On the other hand, the predictions for the scalar spectral index are very similar for the cold and warm power-law plateau inflation models. Regarding the tensor-to-scalar-ratio, in the warm inflation scenario, 
this quantity is suppressed by a factor $(T/H)R^{5/2} > 1$ compared with standard cold inflation. In particular, for $a=1$, the tensor-to-scalar ratio is almost the same order compared to cold power-law plateau inflation. However, for $a=0$ and $a=-1$, the tensor-to-scalar ratio becomes smaller than the cold power-law plateau inflation. 

In a second approach addopted in Ref.\cite{Dimopoulos:2016zhy}, the authors considered a trans-Planckian excursion of the inflaton field, obtaining values for $\alpha$ going from $\alpha=1$ up to $\alpha=5$, and the tensor-to-scalar ratio and the running of the scalar spectral index taking values from $r=0\textup{.}004106$
up to $r=0\textup{.}024412 $, and from $\frac{dn_s}{d \ln k}=-0\textup{.}00057$ up to $\frac{dn_s}{d \ln k}=-0\textup{.}00051$ , respectively. This implies that, for any value of $a$, the tensor-to-scalar ratio for our warm power-law plateau inflation is always lower than the predicted by the cold scenario. On the other hand, the values predicted for the scalar spectral index in the cold and warm scenarios are very similar. In addition, it is interesting to mention that the running of the scalar spectral index $\frac{dn_s}{d\ln k}$
predicted by our warm power-law plateau model is almost  two orders of magnitude greater than the predicted by the cold scenario.

After the analysis performed previously for each value of $a$, we only found a lower limit on $C_{\phi}$ as well as for $\alpha$ and $V_0$, which means that we have a larger range of parameter values to enter in accordance with the Planck results and consistent with a strong dissipative dynamics. This degeneracy could be broken combining these results with the constraints on the inflationary observables related with non-Gaussianities, particularly the $f_{NL}$ parameter, since in warm inflation scenario these have different features when comparing with cold inflation \cite{Bastero-Gil:2014raa}. Despite this issue, the predictions of warm power-law plateau inflation are comparable to those of power-law plateau cold inflation, however the difference between both scenarios is that a way to address the problem of reheating in cold power-law plateau inflation is provided by the warm inflation scenario.

\section{Conclusions}\label{conclu}

In the present work we have studied the consequences of considering 
a new family of single-field inflation models, called power-law plateau inflation, in
the warm inflation scenario. As far we know, this is the first work in studying
the dynamics of warm inflation by using the power-law plateau potential. In order to describe the dissipative effects during
the inflationary expansion, we considered a generalized expression for the inflaton
decay ratio given by $\Gamma(\phi,T)=C_{\phi}\frac{T^a}{{\phi}^{a-1}}$,
where $a=3,1,0,-1$, denotes several inflaton decay ratios studied in the literature.
We restricted ourselves only to study the strong dissipative regime, $R\gg 1$.
For this dissipative regime, under the slow-roll approximation, we 
have studied the background as well as the perturbative dynamics. 
In particular, we have found the expressions for the  scalar power spectrum, scalar spectral index and its running as well as the tensor-to-scalar ratio. Contrary to the standard cold inflation, in the warm inflation scenario it is not
sufficient to consider only the constraints on the $r-n_s$ plane, but we also have to consider
the essential condition for warm inflation $T>H$ and the conditions for the model evolves under strong
dissipative regime $R\gg 1$. For completeness, we study the predictions of our model regarding the running of the scalar spectral index, through the $n_s-dn_s/d \ln k$ plane.

To compare the predictions of power-law plateau inflation in the cold and warm scenarios, we restricted ourselves to the case 
$n=2$ and $q=1$, corresponding to the best choice of model in Ref.\cite{Dimopoulos:2016zhy}. For this particular case, the inflaton decay $a=3$, i.e. $\Gamma \propto \frac{T^3}{\phi^2}$, fails in describe a strong dissipative dynamics consistent with current data, since the predicted value for the scalar spectral index is always greater than unity. We recall that, for the more representative potentials studied in the literature, the inflaton decay rate $a=3$ describes a warm inflationary dynamics consistent with current data. Regarding the predictions in the $n_s-r$ and $n_s-dn_s/d \ln k$ planes, for $a=1$, the tensor to scalar ratio and the running of the spectral index becomes
$r\simeq 10^{-4}$ and $\frac{d n_s}{d\ln k}\simeq 0\textup{.}002$, respectively, whereas for both the cases $a=0$ and $a=-1$, these inflationary observables become $r\simeq 10^{-5}$ and $\frac{d n_s}{d\ln k}\simeq 0\textup{.}001$, being consistent with current bounds imposed by Planck for $\Lambda$CDM $+r+dn_s/d\ln k$. Is interesting to mention that, for other kind of potentials already studied in the warm inflaton scenarios, the decay ratios $a=0$ and $a=-1$ predicted a scalar spectral index always greater than unity. On the other hand, for any value of $a$, the condition for the model evolves according to the strong dissipative regime  sets the lower limit for the disipative parameter $C_{\phi}$  as well for $\alpha$ and $V_0$. However, the essential condition for warm inflation to occur, $T>H$ neither  the Planck data, by considering the two-dimensional marginalized constraints at 68 $\%$ and 95 $\%$ C.L. on the parameters $r$ and $n_s$,  do not impose any constraints on the model for this dissipative regime, obtaining a lower limit on $C_{\phi}$ as well as for $\alpha$ and $V_0$. However, if we consider the
observational constraints on the inflationary observables related with non-Gaussianities, particularly the $f_{NL}$ parameter, this degenerancy in the parameters could be broken. 

Comparing our warm power-law plateau inflation model with the standard one, we found that the strong dissipative warm inflation dynamics is only consistent with a trans-Planckian incursion of the inflaton potential, according with second approach addopted in \cite{Dimopoulos:2016zhy}, ensuring that this power-law plateau potential be distinguishable from monomial inflation. For this trans-Planckian evolution of
the inflaton, and for any value of $a$, the tensor-to-scalar ratio for our warm power-law plateau inflation is always lower than  predicted by the cold scenario. On the other hand, the values predicted for the scalar spectral index in the cold and warm scenarios become similar, however, the running of the scalar spectral index $\frac{dn_s}{d\ln k}$ is almost  two orders of magnitude greater than predicted by the cold scenario. We have shown that warm power-law plateau inflaton, with decay ratios parametrized by $a=1,0$, and $-1$, is consistent with a strong dissipative dynamics and predicts values for the scalar spectral index, the running of the scalar spectral index, and tensor-to-scalar ratio consistent with current bounds imposed by Planck,  for $\Lambda$CDM $+r+dn_s/d\ln k$.

\begin{acknowledgments}
N.V. was supported by Comisi\'on Nacional
de Ciencias y Tecnolog\'ia of Chile through FONDECYT Grant N$^{\textup{o}}$
3150490. Finally, we wish to thank to the anonymous referee for her/his valuable comments, which have helped us to improve the presentation in our manuscript.
\end{acknowledgments}

\end{document}